\documentclass[10pt,conference]{IEEEtran}
\usepackage{cite}
\usepackage{amsmath,amssymb,amsfonts}
\usepackage{graphicx}
\usepackage{textcomp}
\usepackage{xcolor}
\usepackage [english]{babel}
\usepackage [autostyle, english = american]{csquotes}
\MakeOuterQuote{"}
\usepackage{graphicx}
\usepackage{float}
\usepackage{graphics} 
\usepackage{epsfig} 
\usepackage{tabularx}
\usepackage{booktabs}
\usepackage{multicol}
\usepackage{multirow}
\usepackage{tabularx}
\usepackage{pdflscape}
\usepackage{balance}
\usepackage{bbding}
\usepackage{wasysym}
\usepackage{amssymb}
\usepackage{pifont}
\usepackage{amssymb}
\usepackage{afterpage}
\usepackage{subcaption}
\usepackage{capt-of}
\usepackage{bbm}
\usepackage{rotating}
\usepackage{lscape}
\usepackage{hyperref}
\hypersetup{bookmarks=false}
\usepackage{pgfplots}
\pgfplotsset{width=8cm,compat=1.9}
\graphicspath{ {images/} }
\usepackage{amsmath}
\usepackage{braket} 
\usepackage{caption}
\usepackage{algorithm}
\usepackage{xcolor}
\usepackage[noend]{algpseudocode}
\usepackage{mathtools,bm}

\def\BibTeX{{\rm B\kern-.05em{\sc i\kern-.025em b}\kern-.08em
    T\kern-.1667em\lower.7ex\hbox{E}\kern-.125emX}}
    
\makeatletter
\patchcmd{\@maketitle}
  {\addvspace{0.5\baselineskip}\egroup}
  {\addvspace{-1\baselineskip}\egroup}
  {}
  {}
\makeatother

\begin{document}

\title{SplitPlace: Intelligent Placement of Split Neural Nets in Mobile Edge Environments}

\author{
\IEEEauthorblockN{Shreshth Tuli}
\IEEEauthorblockA{\textit{Department of Computing, Imperial College London, United Kingdom} \\
ACM Student Member: 4930025\\
s.tuli20@imperial.ac.uk}
}

\maketitle

\begin{abstract}
In recent years, deep learning models have become ubiquitous in industry and academia alike. Modern deep neural networks can solve one of the most complex problems today, but coming with the price of massive compute and storage requirements. This makes deploying such massive neural networks challenging in the mobile edge computing paradigm, where edge nodes are resource-constrained, hence limiting the input analysis power of such frameworks. Semantic and layer-wise splitting of neural networks for distributed processing show some hope in this direction. However, there are no intelligent algorithms that place such modular splits to edge nodes for optimal performance. This work proposes a novel placement policy, SplitPlace, for the placement of such neural network split fragments on mobile edge hosts for efficient and scalable computing.
\end{abstract}

\begin{IEEEkeywords}
Placement policy; Split Neural Networks; Mobile Edge Computing.
\end{IEEEkeywords}
\section{Problem and Motivation}\label{sec:introduction}
\noindent
Modern Deep Neural Networks (DNN) are becoming the backbone of many industrial tasks and activities~\cite{gill2019transformative}. As the computational capabilities of devices have improved, new deep learning models have been proposed~\cite{zhu2018tbd}. Such neural models are becoming increasingly demanding in terms of data and compute power to provide higher accuracy results in more challenging problems. Many recent DNN models have been shown to outperform the earlier shallow networks for more complex tasks like image segmentation, traffic surveillance, and healthcare~\cite{alom2018history, tuli2020healthfog, tuli2020ithermofog}. Moreover, recently paradigms like mobile edge computing have emerged which provide robust and low latency deployment of Internet of Things (IoT) applications close to the edge of the network.

However, mobile edge devices face the severe limitation of computational and memory resources as they rely on low power energy sources like batteries, solar or other energy scavenging methods~\cite{abbas2017mobile, mao2016dynamic}. This is not only because of the requirement of low cost but also the need for mobility in such nodes~\cite{tuli2020dynamic}. Herein, it is still possible to handle the processing limitations of massive DNN models by effective preemption and longer execution of jobs. However, memory bottlenecks are much harder to solve~\cite{tuli2020dynamic}. In a distributed edge environment where storage spaces a typically mapped to a networks-attached-media, large swap spaces impose very high network bandwidth overheads making high fidelity inference using DNNs hard~\cite{tuli2020healthfog, tuli2021cosco}. To deploy an upgraded AI model, tech-giants like Amazon, Netflix and Google need to revamp their infrastructure and upgrade their devices, raising many sustainability concerns~\cite{gill2019transformative}. This has made the integration of massive neural network models with such devices a challenging and expensive ordeal. 

The only solution for this problem is the development of strategies that can accommodate large-scale DNNs within legacy infrastructures. Many prior efforts in this regard have been proposed~\cite{kim2017splitnet, shi2020communication, lim2020federated} but they fail to provide a holistic strategy for not only distributed learning but also inference in such memory-constrained environments. This abstract proposes to solve these challenges by intelligent splitting and placement of large neural models into modular fragments. This project is part of a larger endeavor to efficiently integrate the hitherto disjoint fields of advanced deep learning and distributed systems research.

\section{Background and Related Work}

Recently, many research ideas have been proposed like Cloud AI, Edge-AI and Federated learning which aim to solve the problem of running enormous deep learning models on constrained edge devices~\cite{shi2020communication, lim2020federated}. However, Cloud-AI faces the problem of high average response time making it unsuitable for latency-critical applications like Healthcare~\cite{tuli2020healthfog, tuli2020shared}. Edge-AI requires all training data to be centralized leading to high bandwidth overheads~\cite{shi2020communication}. Federated learning depends on data distribution over multiple nodes and assumes that neural models with data batches can be accommodated in the system memory. This is seldom the case for low-end nodes like Raspberry Pis~\cite{chen2019deep}.

Other recent works propose lower precision models that can fit in the limited memory of such devices by using methods like Model Compression~\cite{capotondi2020cmix, gunasekaran2020implications}. However, compressed and low-precision models lose the inference accuracy making them unsuitable for accuracy sensitive applications like security and intrusion detection~\cite{le2020overview}. Recently, split neural network models have been proposed which show that using semantic or layer-wise split, a large deep neural network can be split to multiple smaller networks for dividing network parameters into multiple nodes \cite{matsubara2019distilled, ushakov2018split, kim2017splitnet, gordon2008self}. However, no appropriate scheduling policies exist which can intelligently place such modular neural fragments on a distributed infrastructure to optimize both accuracy and Service Level Agreements (SLA). 

\begin{figure*}[!t]
    \centering
    \begin{subfigure}{\columnwidth}
        \centering
        \includegraphics[width=0.8\linewidth]{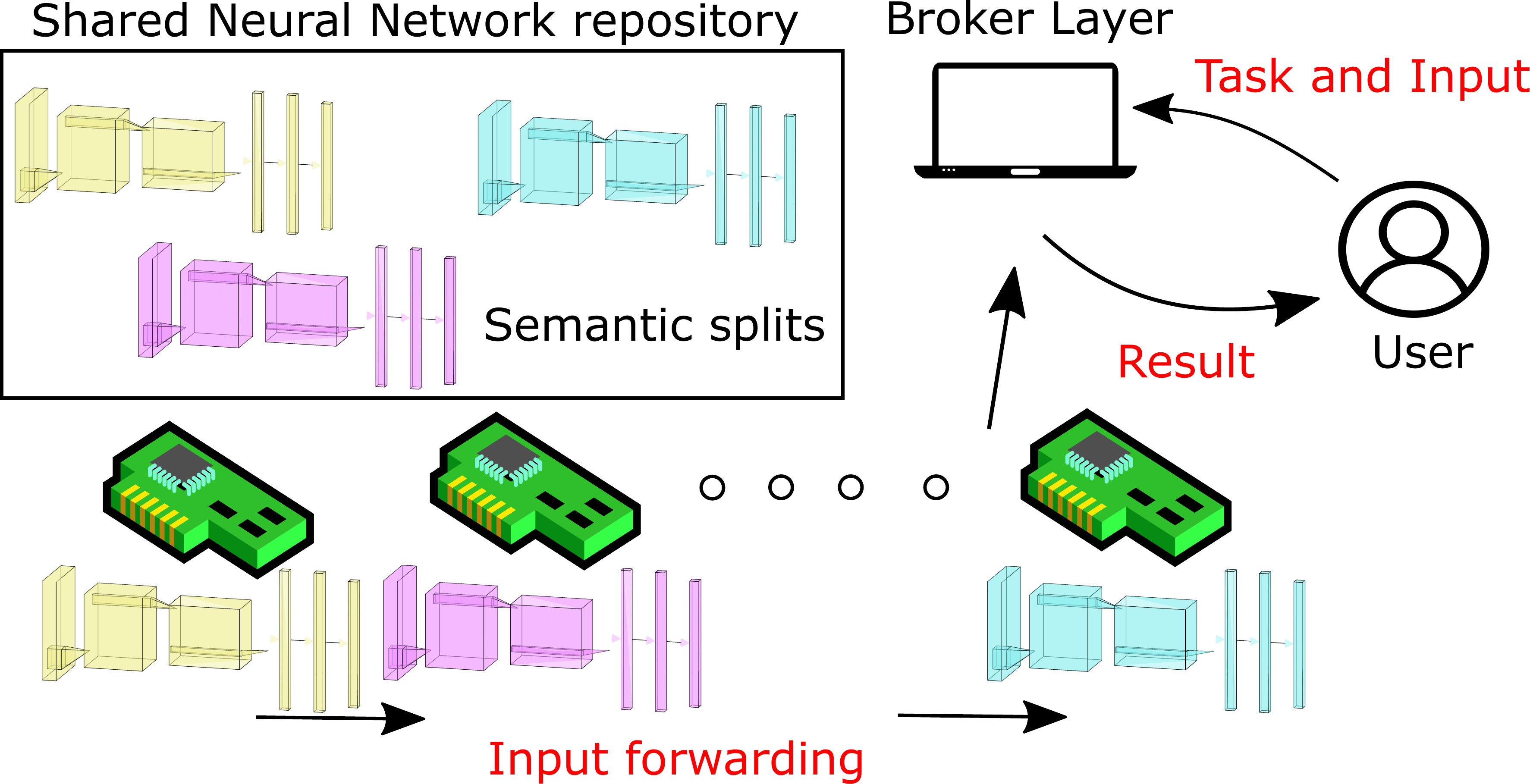}
        \caption{Semantic split execution}
        \label{fig:sem}
    \end{subfigure}
    \begin{subfigure}{\columnwidth}
        \centering
        \includegraphics[width=0.8\linewidth]{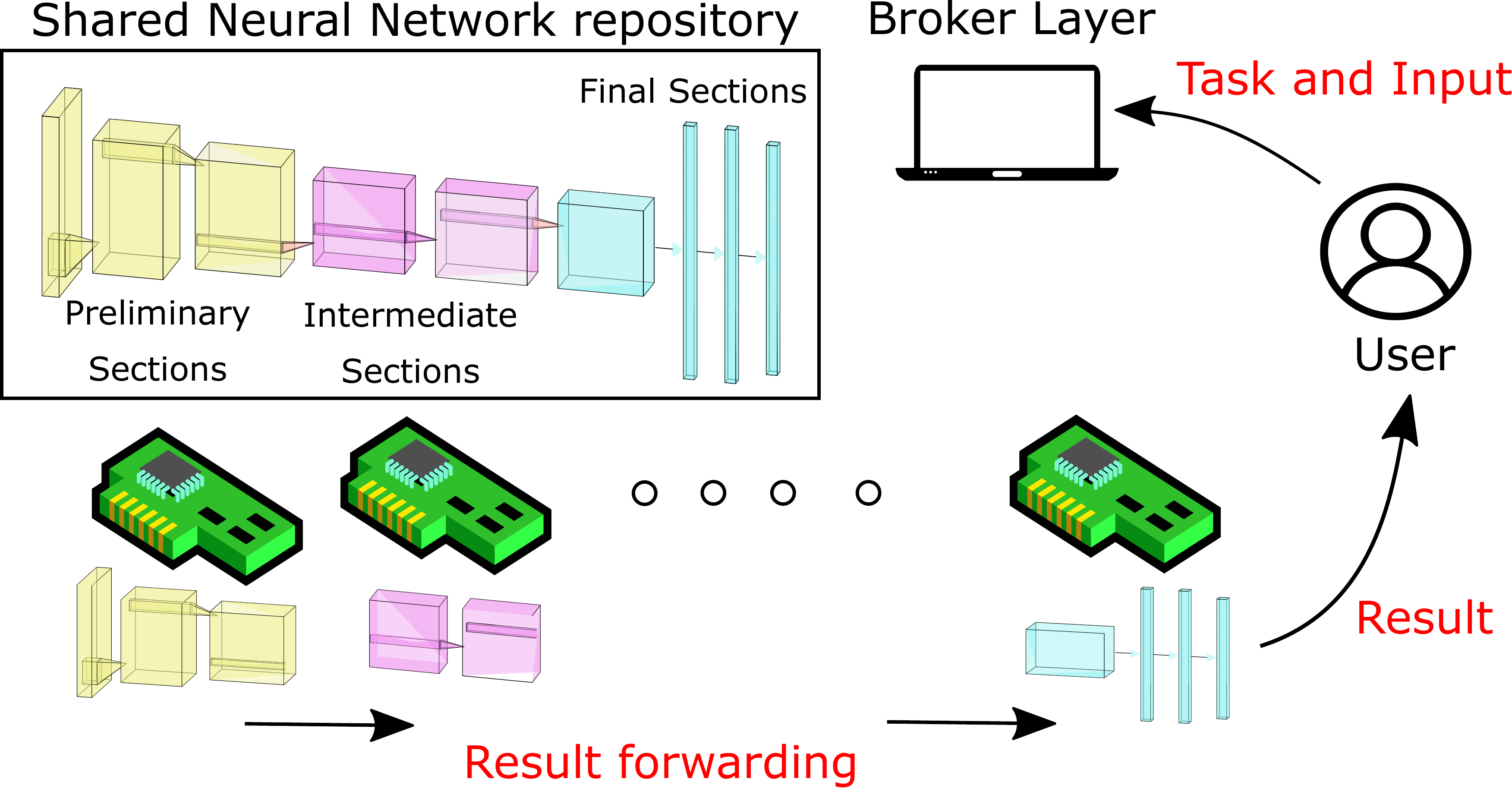}
        \caption{Layer split execution}
        \label{fig:layer}
    \end{subfigure}
    \caption{Proposed scalable deep learning architecture}
    \label{fig:scale}
\end{figure*}

\section{Proposed Method}

\subsection{Layer and Semantic Splits}

Semantic splitting divides the network weights into a hierarchy of multiple groups that use a different set of features which produces a tree structured model which has no connection among branches of the tree allowing parallelization of input analysis. Due to limited information sharing among neural network fragments, the semantic splitting scheme gives lower accuracy in general. Semantic splitting requires a separate training procedure where pre-trained models can not be used, however, it provides parallel task processing  and hence lower inference times more suitable for mission-critical tasks like healthcare and surveillance. Layer wise splitting divides the network into groups of layers for sequential processing of the task input. Layer splitting is easier to deploy as pre-trained models can be just divided into multiple layer groups and distributed to different mobile edge nodes. However, layer splits require semi-processed input to be forwarded to subsequent edge node with final processed output to be sent to the user, thus increasing the overall execution times. Moreover, layer-wise splitting gives higher accuracy compared to semantic splitting. An overview of these two strategies is shown in Figure~\ref{fig:scale}.
 
\subsection{Intelligent Placement}
However, the placement of such split models is non-trivial considering the diverse and complex dynamism of task distribution, model usage frequencies and geographical placement of mobile edge devices~\cite{ahmed2017mobile}. This work proposes a novel split placement policy, \textbf{SplitPlace}, for enhanced computation at the edge of the leveraging mobile edge platform to reach low latency results as well as allowing modular neural models to be integrated for best result accuracies that only cloud deployments could provide till date. The idea behind the proposed placement policy is to maintain moving average estimates of the time it takes for the complete execution of the "layer" split decision ($E_a$ for application $a\in A$). Based on the SLA deadline of a new workload, we run two Multi-Armed-Bandit (MAB) models to estimate the "expected reward" of each decision. Motivation being that SLA deadline being lower than the execution time of a layer split, such a decision would more likely lead to a SLA violation. The reward for a sequence of decisions on a set of workloads ($W$) is expressed as
\[\frac{\sum_{w \in W} \left[\mathbbm{1}(\text{Response Time}_w \leq SLA_w) + \text{Accuracy}_w \right]}{2 \cdot |W|},\]
where $SLA_w$ is the deadline of workload $w$. This method is agnostic to the underlying scheduling strategy. However, for comparison with baselines, we combine it with a popular Asynchronous-Actor-Critic scheduler~\cite{tuli2020dynamic}.

\begin{figure}[!t]
    \centering
    \includegraphics[width=0.8\linewidth]{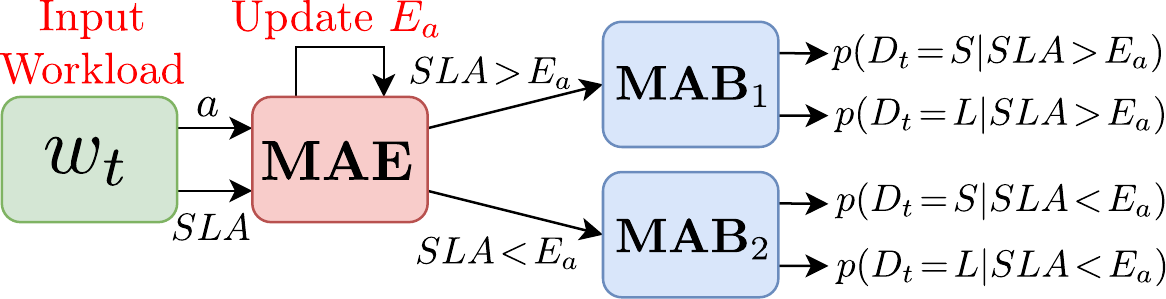}
    \caption{Proposed Split Decision Model.}
    \label{fig:decider}
\end{figure}

These MAB models are learned and combined with an intelligent decision-aware scheduler for the dynamic placement of DNN workloads on a mobile edge computing environment. An overview of the proposed method is shown in Figure~\ref{fig:decider} where $D_t$ is the decision (layer or semantic splitting) for workload $w_t$ received at time $t$ with application type $a\in A$. Fine-grained decision making on how to optimally split DNNs can be explored in the future.

\section{Performance Evaluation}

To compare the proposed approach, we use a resource-constrained edge setup with 10 Raspberry-Pi like devices having 4 to 8 GB of RAM available per device. To emulate mobility, we use Gaussian noise in the network latency using the \textit{netlimiter} Linux tool~\cite{tuli2020dynamic}. We use a popular image-classification models of ResNet50-V2, MobileNetV2 and InceptionV3~\cite{gunasekaran2020implications}. We use popular model-compression strategy as a baseline giving low memory-footprint versions of this model to run on the test setup as vanilla models can not be run directly in such constrained devices~\cite{teerapittayanon2017distributed, gunasekaran2020implications}.

Table~\ref{tab:comparison} shows the results of our experiments. With a $10.6\%$ higher scheduling time compared to baseline, the SplitPlace model is able to give lower energy consumption and SLA violation by $5\%$ and $61\%$ respectively. Moreover, the proposed model gives nearly $3\%$ higher inference accuracy and $9.6\%$ higher average reward compared to the baseline.

\begin{table}[]
    \centering
    \caption{Comparison with the baseline.}
    \resizebox{\linewidth}{!}{
    \begin{tabular}{@{}lccccc@{}}
    \toprule 
    \textbf{Model} & \textbf{Energy} & \textbf{Sched. Time} & \textbf{SLA violation} & \textbf{Accuracy} & \textbf{Reward}\tabularnewline
    \midrule
    Baseline & 94.88 & \textbf{4.42$\pm$0.02} & 0.21$\pm$0.02 & 89.93\% & 83.98\%\tabularnewline
    SplitPlace & \textbf{90.12} & 4.89$\pm$0.09 & \textbf{0.08$\pm$0.02} & \textbf{91.07\%} & \textbf{90.11\%}\tabularnewline
    \bottomrule
    \end{tabular}
    }
    \label{tab:comparison}
\end{table}

\section*{Acknowledgments}
\footnotesize{The author is supported by the President’s Ph.D. Scholarship at the Imperial College London. This project is part of the author's Ph.D. work, co-supervised by Dr. Giuliano Casale and Prof. Nick Jennings.}

\balance
\bibliographystyle{IEEEtran}

\bibliography{references}

\end{document}